\documentclass[conference]{IEEEtran}

\usepackage{amsmath}
\usepackage{amssymb}
\usepackage{latexsym}
\usepackage{array,arydshln}
\usepackage{multirow}
\usepackage{graphicx}
\usepackage{float}
\usepackage{bm}
\usepackage{breqn}
\usepackage{url}
\usepackage{ctable}

\newtheorem{theorem}{Theorem}
\newtheorem{corollary}{Corollary}

\newcommand{\bs}[1]{\ensuremath{\boldsymbol{#1}}}

\newcommand{\beq}{\begin{equation}}
\newcommand{\enq}{\end{equation}}
\newcommand{\beal}{\begin{align*}}
\newcommand{\enal}{\end{align*}}

\IEEEoverridecommandlockouts

\begin{document}

\title{Finite Length Weight Enumerator Analysis of Braided Convolutional Codes}


\author{
\IEEEauthorblockN{Saeedeh Moloudi$^\dag$, Michael Lentmaier$^\dag$, and Alexandre Graell i Amat$^\ddag$}
\IEEEauthorblockA{$\dag$Department of Electrical and Information
  Technology, Lund University, Lund, Sweden \\
  $\ddag$Department of Signals and Systems, Chalmers University of Technology, Gothenburg, Sweden\\
              \{saeedeh.moloudi,michael.lentmaier\}@eit.lth.se, alexandre.graell@chalmers.se}\\
              \thanks{This work was supported in part by the Swedish Research Council (VR) under grant \#621-2013-5477.}\vspace*{-1cm}
}


\maketitle


\begin{abstract}
Braided convolutional codes (BCCs) are a class of spatially coupled turbo-like codes (SC-TCs) with excellent belief propagation (BP) thresholds. 
In this paper we analyze the performance of BCCs in the finite block-length regime.
We derive the average weight enumerator function (WEF) and compute the union bound on the performance for the uncoupled BCC ensemble.
Our results suggest that the union bound is affected by poor distance properties of a small fraction of codes. 
By computing the union bound for the expurgated ensemble, we show that the floor improves substantially and very low error rates can be achieved for moderate permutation sizes.
 Based on the WEF, we also obtain a bound on the minimum distance which indicates that it grows linearly with the permutation size.
Finally, we show that the estimated error floor for the uncoupled BCC ensemble is also valid for the coupled ensemble by proving that the minimum distance
of the coupled ensemble is lower bounded by the minimum distance of the uncoupled ensemble.
\end{abstract}

\IEEEpeerreviewmaketitle

\section{Introduction}

Low-density parity-check (LDPC) convolutional codes \cite{JimenezLDPCCC}, also known as spatially coupled LDPC (SC-LDPC) codes \cite{Kudekar_ThresholdSaturation}, have attracted a lot of attention because they exhibit a threshold saturation phenomenon: the belief propagation (BP) decoder can achieve the threshold of the optimal maximum-a-posteriori (MAP) decoder. 
Spatial coupling is a general concept that is not limited to LDPC codes.
Spatially coupled turbo-like codes (SC-TCs) are proposed in \cite{Moloudi_SCTurbo}, where some block-wise spatially coupled ensembles of parallel concatenated codes (SC-PCCs)
and serially concatenated codes (SC-SCCs) are introduced. 
Braided convolutional codes (BCCs) \cite{ZhangBCC} are another class of SC-TCs.
The original BCC ensemble has an inherent spatially coupled structure with coupling memory $m=1$. 
Two extensions of BCCs to higher coupling memory, referred to as Type-I and Type-II BCCs, are proposed in \cite{Moloudi_SPCOM14}.

The asymptotic behavior of  BCCs,
SC-PCCs and SC-SCCs is analyzed in \cite{JournalMLD} where the exact density evolution (DE) equations are derived for the binary erasure channel (BEC). 
Using DE, the thresholds of the BP decoder are computed for both uncoupled and coupled ensembles and compared with the corresponding MAP thresholds.
The obtained numerical results demonstrate that threshold saturation occurs for all considered SC-TC ensembles if the coupling memory is large enough.
Moreover, the occurrence of threshold saturation is proved analytically for SC-TCs over the BEC in \cite{JournalMLD,Moloudi_ITW15}.
While the uncoupled BCC ensemble suffers from a poor BP threshold,
the BP threshold of the coupled ensemble improves significantly even for
coupling memory $m=1$. 
Comparing the BP thresholds of SC-TCs in \cite{JournalMLD}
indicates that for a given coupling memory, the Type-II BCC ensemble
has the best BP threshold for almost all code rates. 

Motivated by the good asymptotic performance and the excellent BP thresholds of BCCs, our aim in this paper is analyzing the performance of BCCs in the
finite block-length regime by means of the ensemble weight enumerator.  
As a first step, we derive the finite block-length
ensemble weight enumerator function (WEF) of the uncoupled ensemble by considering uniform random permutations.
Then we compute the union bound for uncoupled BCCs. 
The unexpectedly high error floor predicted by the bound suggests that the bound is affected by the bad performance of codes with poor distance properties.
We therefore compute the union bound on the performance of the expurgated ensemble by excluding the codes with poor distance properties. 
The expurgated bound demonstrates very low error floors for moderate permutation sizes. 
We also obtain a bound on the minimum distance of the BCC ensemble which reveals that the minimum distance grows linearly with the permutation size.  
 
Finally, we prove that the codeword weights of the coupled ensemble are lower bounded by those of the uncoupled ensemble. 
Thus, the minimum distance of the coupled BCC ensemble is larger than the minimum distance of the uncoupled BCC ensemble. 
From this, we conclude that the estimated error floor of the uncoupled ensemble is also valid for the coupled ensemble.

\section{Compact Graph Representation of \\ Turbo-Like Codes}
In this section, we describe three ensembles of turbo-like codes, namely PCCs, uncoupled BCCs, and coupled BCCs, using the compact graph representation introduced in \cite{JournalMLD}.
 \subsection{Parallel Concatenated Codes}
Fig.~\ref{CGPCCBCC}(a) shows the compact graph representation of a PCC ensemble
with rate $R=\frac{N}{3N}=\frac{1}{3}$, where $N$ is the permutation size. 
These codes are built of two rate-$1/2$ recursive systematic convolutional encoders, referred to as the upper and lower component encoder. 
The corresponding trellises are denoted by $\text{T}^{\text{U}}$ and $\text{T}^{\text{L}}$, respectively. 
In the graph, factor nodes, represented by squares,  correspond to trellises.
All
information and parity sequences are shown by black circles, called variable nodes.
The information sequence, $\bs{u}$, is
connected to factor node $\text{T}^{\text{U}}$ to produce the upper parity
sequence $\bs{v}^{\text{U}}$. Similarly, a reordered copy of $\bs{u}$
is connected to $\text{T}^{\text{L}}$ to produce $\bs{v}^{\text{U}}$.  In order to emphasize that a reordered
copy of $\bs{u}$ is used in $\text{T}^{\text{L}}$, the permutation
is depicted by a line that crosses the edge which connects $\bs{u}$
to $\text{T}^{\text{L}}$. 

\subsection{Braided Convolutional Codes}
\subsubsection{Uncoupled BCCs}
\begin{figure}[t]
  \centering
    \includegraphics[width=0.5\linewidth]{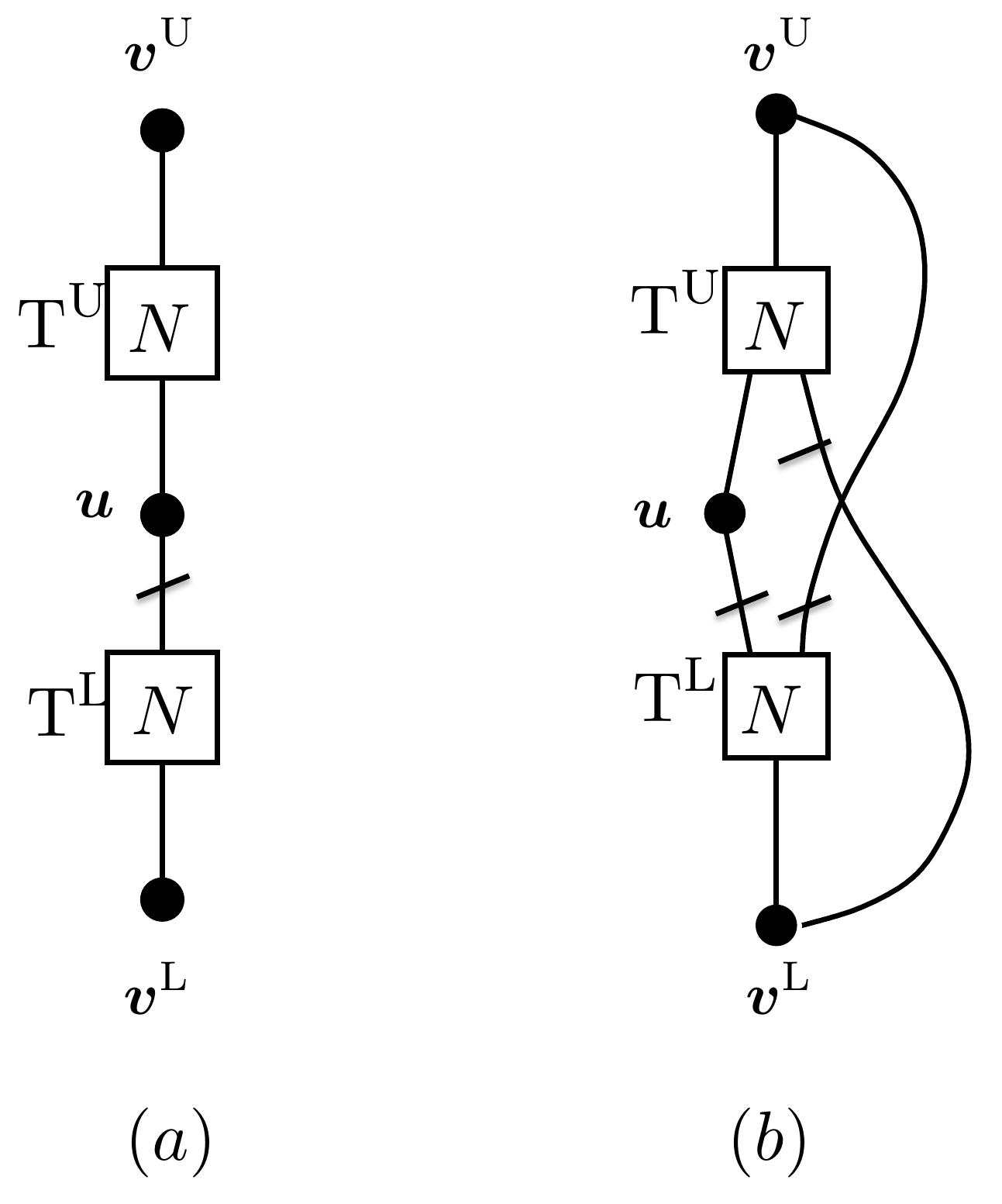}
\caption{(a) Compact graph representation of (a) PCCs  (b) BCCs.}
\vspace{-2mm}
\label{CGPCCBCC}
\vspace{-3mm}
\end{figure}

The original BCCs are inherently a class of SC-TCs \cite{Moloudi_SCTurbo,ZhangBCC,Moloudi_SPCOM14}. 
An uncoupled BCC ensemble can be obtained by tailbiting a BCC ensemble with coupling length $L=1$.
The compact graph representation of this ensemble is shown in Fig.~\ref{CGPCCBCC}(b).
The BCCs of rate $R=\frac{1}{3}$ are built of two rate-$2/3$
recursive systematic convolutional encoders. The corresponding
trellises are denoted by $\text{T}^{\text{U}}$ and
$\text{T}^{\text{L}}$, and referred to as the upper and lower trellises,
respectively.
The information sequence $\bs{u}$ and a reordered version of the
lower parity sequence
$\bs{v}^{\text{L}}$ are connected to $\text{T}^{\text{U}}$ to produce
the upper parity sequence $\bs{v}^{\text{U}}$. Likewise, a
reordered version of $\bs{u}$ and a reordered version of
$\bs{v}^{\text{U}}$ are connected to $\text{T}^{\text{L}}$ to produce
$\bs{v}^{\text{L}}$.

\subsubsection{Coupled BCCs, Type-I}
Fig.~\ref{SCBCC}(a) shows the compact graph representation of the original
BCC ensemble, which can be classified as Type-I BCC ensemble
\cite{Moloudi_SPCOM14} with coupling memory $m=1$. As depicted in
Fig.~\ref{SCBCC}(a), at time $t$, the information sequence $\bs{u}_{t}$ and a reordered version of the lower
parity sequence at time $t-1$, $\bs{v}_{t-1}^{\text{L}}$, are
connected to $\text{T}_{t}^{\text{U}}$ to produce the current upper parity
sequence $\bs{v}_{t}^{\text{U}}$. Likewise, a reordered version of $\bs{u}_t$ and
$\bs{v}_{t-1}^{\text{U}}$ are connected to $\text{T}_{t}^{\text{L}}$
to produce $\bs{v}_{t}^{\text{L}}$. At time $t$, the inputs of
the encoders come only from time $t$ and $t-1$,
hence the coupling memory is  $m=1$.

\subsubsection{Coupled BCCs, Type-II}
Fig.~\ref{SCBCC}(b) shows the compact graph representation
of Type-II BCCs with coupling memory $m=1$.
As depicted in the figure, in addition to the coupling of the parity sequences,
the information sequence is also coupled. At time $t$, the information
sequence $\bs{u}_t$ is divided into two sequences $\bs{u}_{t,0}$ and
$\bs{u}_{t,1}$.
Likewise, a reordered copy of the information sequence,
$\bs{\tilde{u}}_t$, is divided into two sequences $\bs{u}_{t,0}$ and
$\bs{u}_{t,1}$. At time $t$, the first inputs of the upper and lower
encoders are reordered versions of the sequences
$(\bs{u}_{t,0},\bs{u}_{t-1,1})$ and
$(\tilde{\bs{u}}_{t,0},\tilde{\bs{u}}_{t-1,1})$, respectively. 
    
\section{Input-Parity Weight Enumerator}
\subsection{Input-Parity Weight Enumerator for Convolutional Codes}
Consider a rate-$2/3$ recursive systematic convolutional encoder.
The input-parity weight enumerator function (IP-WEF), $A(I_1,I_2,P)$, can be written as
\[
A(I_1,I_2,P)=\sum_{i_1} \sum_{i_2}\sum_{p} A_{i_1,i_2,p}I^{i_1}I^{i_2}P^p,
\]
where $A_{i_1,i_2,p}$ is the number of codewords with weights $i_1$, $i_2$, and $p$ for the
first input, the second input, and the parity sequence, respectively. 

To compute the IP-WEF, we can define a transition matrix between
trellis sections denoted by $\bs{M}$.
This matrix is a square matrix whose element in the $r$th row and the $c$th
column $[\bs{M}]_{r,c}$ corresponds to the trellis branch
which starts from the 
$r$th state and ends up at the $c$th state. More precisely,
 $[\bs{M}]_{r,c}$ is a monomial
$I_1^{i_1}I_2^{i_2}P^{p}$, where $i_1$, $i_2$, and $p$ can be zero or
one depending on the branch weights.

\begin{figure}[t]
  \centering
    \includegraphics[width=\linewidth]{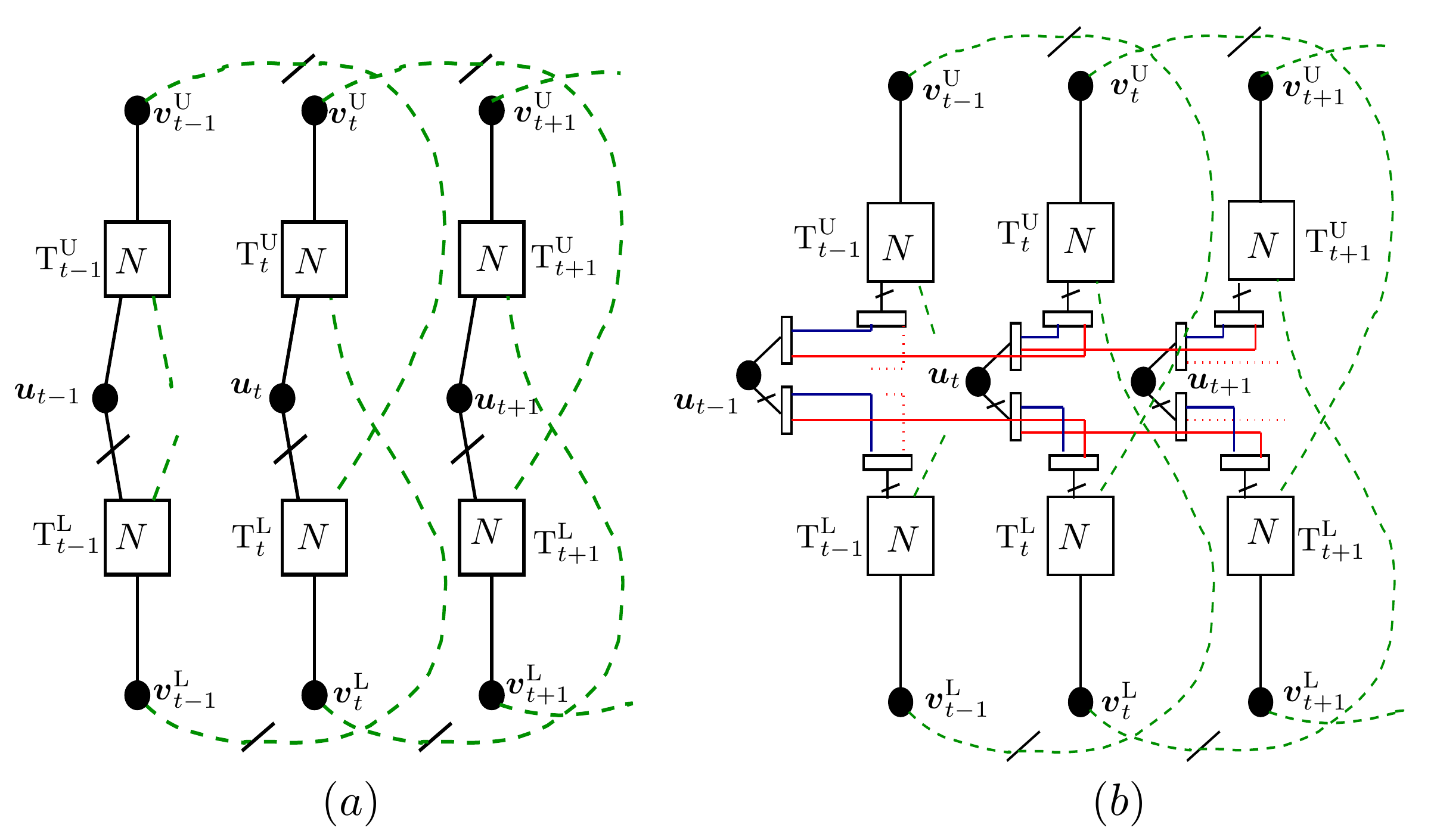}
\vspace{-2mm}
\caption{Compact graph representation of coupled BCCs with coupling
  memory  $m=1$ (a) Type-I (b) Type-II. }
\label{SCBCC}
\vspace{-3mm}
\end{figure}

For a rate-$2/3$ convolutional encoder with generator matrix 
\begin{equation}
\label{BCCG}
\boldsymbol{G}= \left( \begin{array}{ccc}1&0&1/7\\0&1&5/7\end{array}\right),
\end{equation} 
in octal notation, the matrix $\bs{M}$ is
\begin{equation}
\label{eq:AR23}
\boldsymbol{M}(I_1,I_2,P)=\left( \begin{array}{cccc}1&I_2P&I_1I_2&I_1P\\I_1&I_1I_2P&I_2&P\\I_2P&1&I_1P&I_1I_2\\I_1I_2P&I_1&P&I_2 \end{array}\right).
\end{equation}

Assume termination of the encoder after $N$ trellis sections. The
IP-WEF can be obtained by computing $\bs{M}^{N}$.
The element $[\bs{M}^{N}]_{1,1}$  of the resulting matrix 
is the corresponding IP-WEF.

The WEF of the encoder is defined as
\[
A(W)=\sum_{w=1}^{N}A_{w}W^w=A(I_1,I_2,P)\vert_{I_1=I_2=P=W},
\]
where $A_w$ is the number of codewords of weight $w$. 

In a similar way, we can obtain the matrix $\bs{M}$ for a rate-$1/2$
convolutional encoder. The IP-WEF of the encoder is 
$[\bs{M}^{N}]_{1,1}$ and is given by
\[
A(I,P)=\sum_{i} \sum_{p}A_{i,p} I^iP^p,
\]
where $A_{i,p}$ is the number of codewords of input weight $i$ and
parity weight $p$.
\subsection{Parallel Concatenated Codes}
For the PCC ensemble in Fig.~\ref{CGPCCBCC}(a), the IP-WEFs of the upper and lower encoders are defined by
$A^{\text{T}_{\text{U}}}(I,P)$ and $A^{\text{T}_{\text{L}}}(I,P)$, respectively.
The IP-WEF of the overall encoder, $A^{\text{PCC}}(I,P)$, depends on the
permutation that is used, but we can compute the
average IP-WEF for the ensemble.
The coefficients of the IP-WEF of the PCC ensemble \cite{UnveilingTC} can be written as
\begin{equation}
\bar{A}_{i,p}^{\text{PCC}}=\frac{\sum_{p_1}A^{\text{T}_{\text{U}}}_{i,p_1}\cdot A^{\text{T}_{\text{L}}}_{i,p-p_1}}{\binom{N}{i}}.
\end{equation}
\subsection{Braided Convolutional Codes}
For the uncoupled BCC ensemble depicted in
Fig.~\ref{CGPCCBCC}(b), the IP-WEFs of the upper and
lower encoders are denoted by $A^{\text{T}_{\text{U}}}(I_1,I_2,P)$
and $A^{\text{T}_{\text{L}}}(I_1,I_2,P)$, respectively. To derive the average WEF, we
have to average over all possible combinations of permutations.
The coefficients of the IP-WEF of the uncoupled BCC ensemble can be written as 
\begin{equation}
\label{IPWEBCC}
\bar{A}_{i,p}^{\text{BCC}}=\frac{\sum_{p_1}A^{\text{T}_{\text{U}}}_{i,p_1,p-p_1}\cdot A^{\text{T}_{\text{L}}}_{i,p-p_1,p_1}}{\binom{N}{i}\binom{N}{p_1}\binom{N}{p-p_1}}.
\end{equation}
{\it{Remark:}}
It is possible to interpret the BCCs in Fig.~\ref{CGPCCBCC}(b) as
protograph-based generalized LDPC codes with trellis constraints. As a
consequence, the IP-WEF of the ensemble can also be computed by the
method presented in \cite{AbuSurrahGlobeCom07, AbuSurra2011}.

\section{Performance Bounds for Braided Convolutional Codes}
\subsection{Bounds on the Error Probability}
 Consider the PCC and BCC ensembles in Fig.~\ref{CGPCCBCC} with
 permutation size $N$. For transmission over an additive white Gaussian noise (AWGN) channel, the bit error rate (BER) of the code is
upper bounded by
\begin{equation}
\label{BER}
P_b\leq \sum_{i=1}^{N} \sum_{p=1}^{2N} \frac{i}{N} \bar{A}_{i,p}
Q\left ( \sqrt{2(i+p)R\frac{E_b}{N_0}}\right),
\end{equation}
and the frame error rate (FER) is upper bounded by
\begin{equation}
\label{FER}
P_F\leq \sum_{i=1}^{N} \sum_{w=1}^{2N} \bar{A}_{i,p} Q\left( \sqrt{2(i+p)R\frac{E_b}{N_0}}\right),
\end{equation}
where $Q(.)$ is the $Q$-function and $\frac{E_b}{N_0}$ is the signal-to-noise ratio.

\begin{figure}[t]
  \centering
    \includegraphics[width=0.9\linewidth]{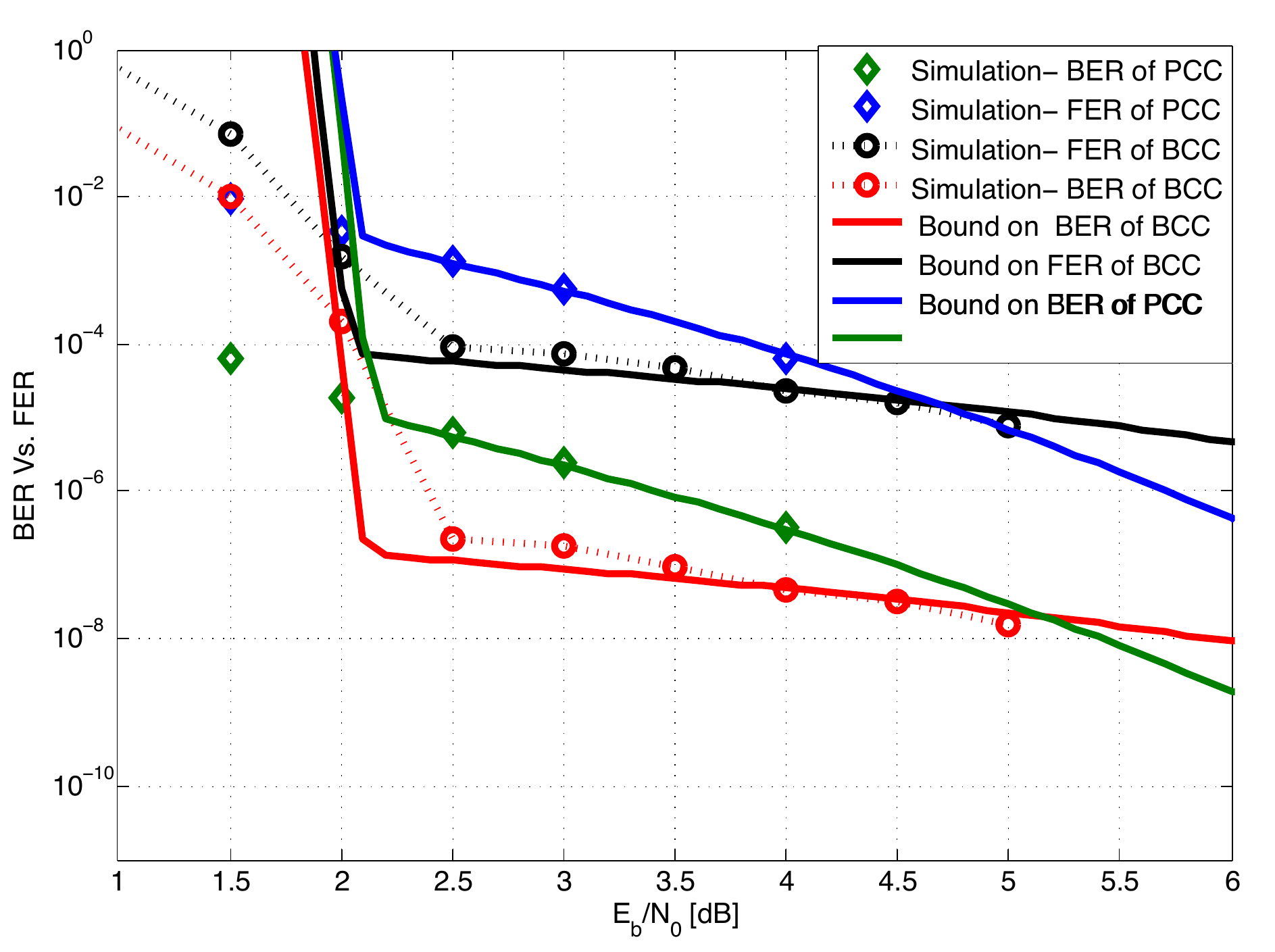}
\vspace{-2mm}
\caption{Simulation results and bound on performance of the PCC and BCC ensembles.}
\label{PCCBCC}
\vspace{-3mm}
\end{figure}
The truncated union bounds on the BER and FER of the PCC ensemble in
Fig.~\ref{CGPCCBCC}(a) are shown in Fig.~\ref{PCCBCC}.  We have considered identical
component encoders with generator matrix $\bs{G}=(1,5/7)$ in octal
notation and permutation size $N=512$.
We also plot the bounds for the uncoupled BCC ensemble with
identical component encoders with generator matrix given in
\eqref{BCCG}. The bounds are truncated at a
value greater than the corresponding Gilbert-Varshamov bound. 
 
\begin{figure}[t]
  \centering
    \includegraphics[width=0.9\linewidth]{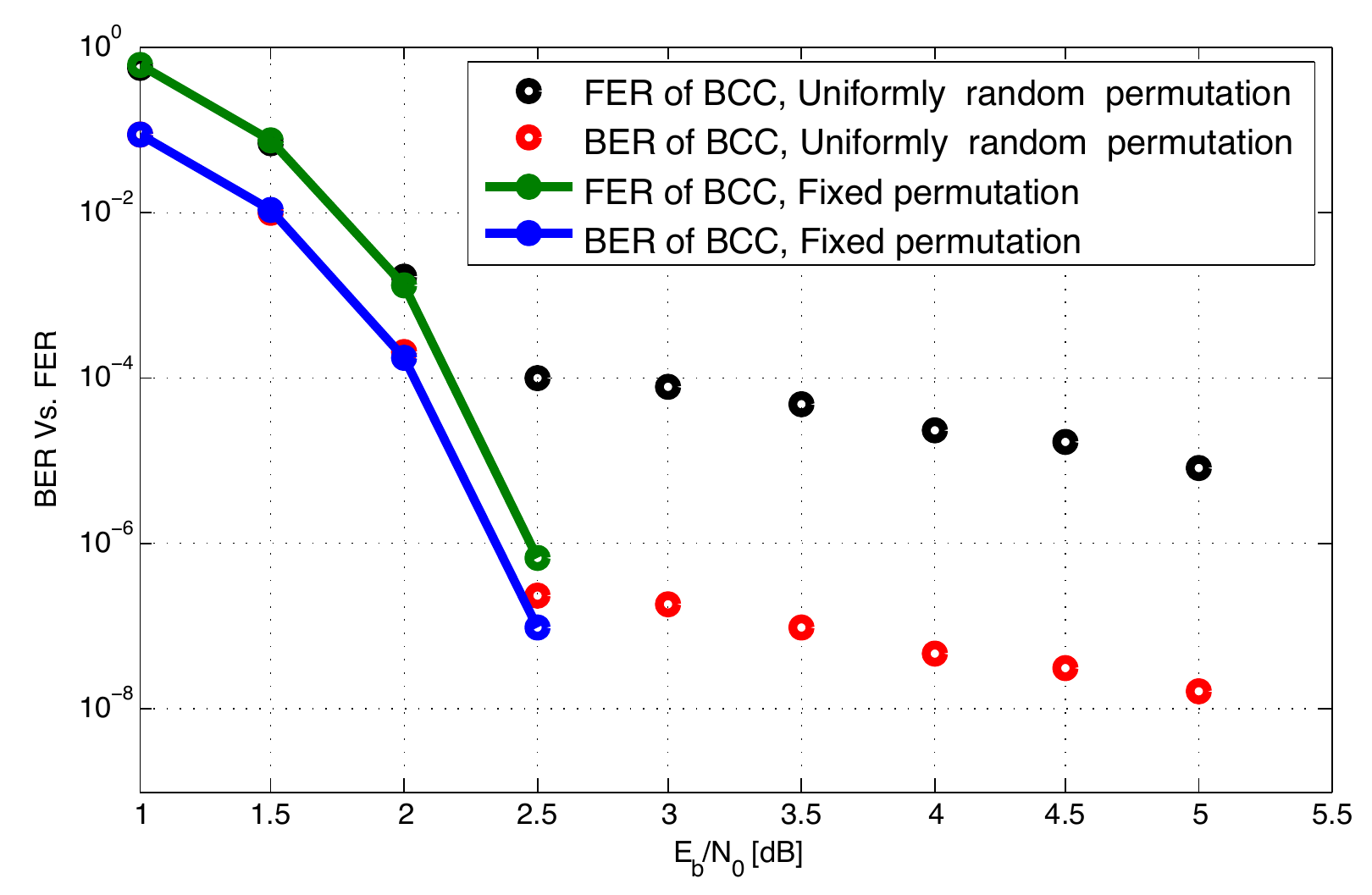}
\vspace{-2mm}
\caption{Simulation results for BCC with uniformly random permutations
  and fixed permutations.}
\label{BCCFP}
\vspace{-3mm}
\end{figure}
Simulation results for the PCC and the uncoupled BCC ensemble are also provided in
Fig.~\ref{PCCBCC}. To simulate the average performance, we have
randomly selected new permutations for each simulated block.
The simulation results are in agreement with the bounds for both ensembles.
It is interesting to see that the error floor for the BCC ensemble is
quite high and the slope of the floor is even worse than that of
the PCC ensemble.  

Fig.~\ref{BCCFP} shows simulation results for BCCs with randomly
selected but fixed permutations.  According to the figure, for
the BCC with fixed permutations, the
performance improves and no error floor is observed.
For example, at $\frac{E_b}{N_0}=2.5\text{dB}$, the FER improves from
$9.5\cdot 10^{-5}$ to $6.8\cdot 10^{-7}$.
Comparing the simulation results for permutations selected uniformly
at random and
fixed permutations, suggests that the bad performance of the
BCC ensemble is caused by a fraction of codes with poor distance
properties. 
In the next subsection, we demonstrate that the performance of  BCCs
improves significantly if we use expurgation.
\subsection{Bound on the Minimum Distance and Expurgated Union Bound} 
Using the average WEF, we can derive a bound on the minimum
distance. We assume that all codes in the ensemble are selected with equal
probability. Therefore, the total number of codewords of weight $w$
over all codes in the ensemble is  $N_{c}\cdot \bar{A}_w$, where $N_c$ is the number of possible
 codes. As an example, $N_c$ is equal to $(N!)^3$ for the BCC
 ensemble.

Assume that
 \begin{equation}
\label{EX}
\sum^{\hat{d}-1}_{w=1}\bar{A}_w<1-\alpha,
\end{equation}
for some integer value $\hat{d}>1$ and a given $\alpha$, $0\leq\alpha<1$.
Then a fraction $\alpha$ of the codes cannot contain codewords of
weight $w<\hat{d}$. If we exclude the remaining fraction $1-\alpha$ of
codes with poor distance properties, the minimum distance of the
remaining codes is lower bounded by $d_{\min}\geq \hat{d}$. 

\begin{figure}[t]
  \centering
    \includegraphics[width=0.9\linewidth]{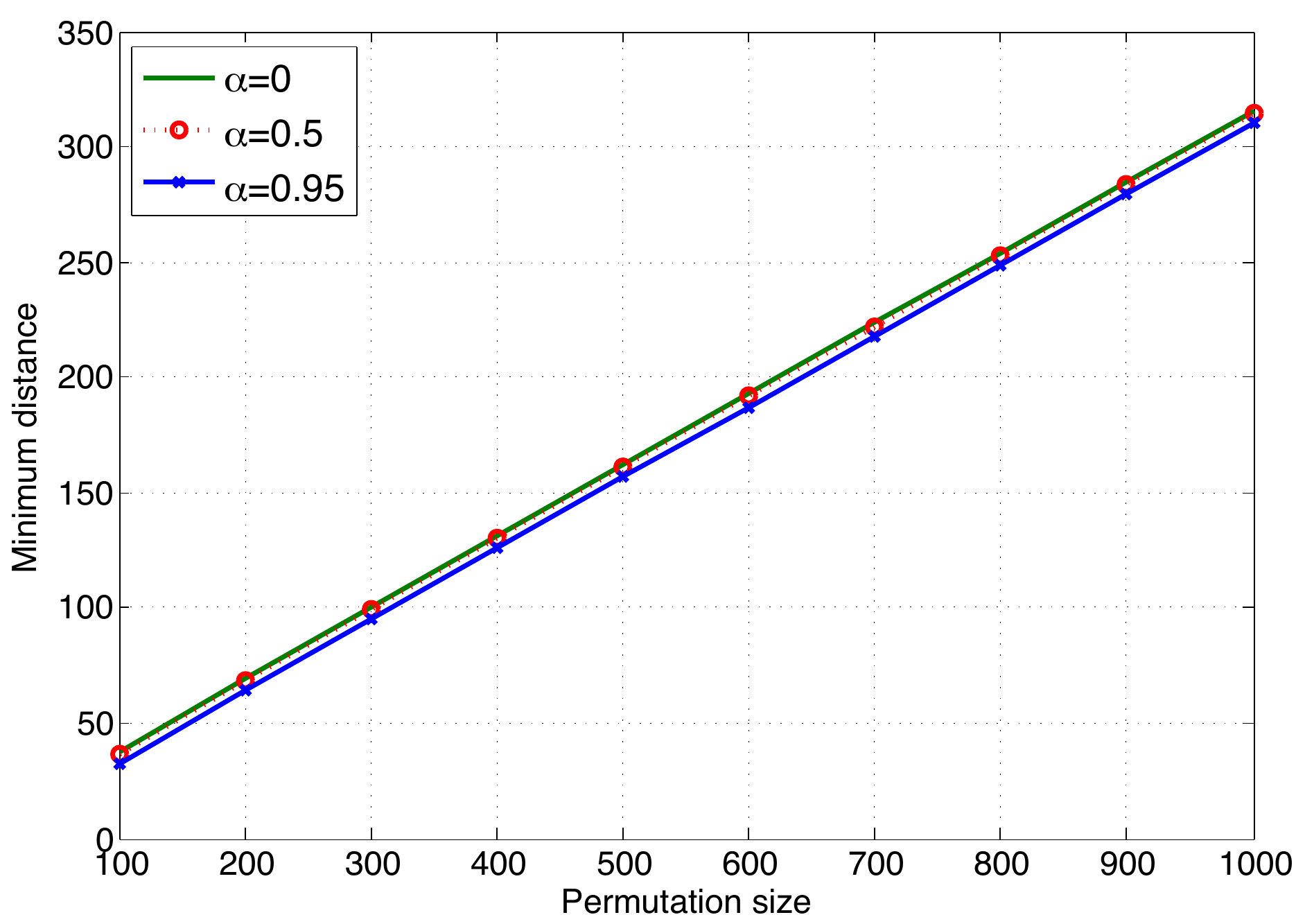}
\vspace{-2mm}
\caption{Bound on the minimum distance for the BCC ensemble.}
\label{MinDis}
\vspace{-3mm}
\end{figure}

The best bound can be obtained by computing the largest $\hat{d}$ that satisfies the condition in \eqref{EX}. 
Considering $\bar{A}_w$ for different permutation sizes, this bound is shown in Fig.~\ref{MinDis} for $\alpha=0$, $\alpha=0.5$, and $\alpha=0.95$. 
According to the figure, the minimum distance of the BCC ensemble grows linearly with the permutation size. The bound corresponding to
$\alpha=0.95$, which is obtained by excluding only $5\%$ of the codes, is very
close to the existence bound for $\alpha=0$.
 This means that only a small fraction of the permutations leads to poor distance properties. 

Excluding the codes with $d_{\min}<\hat{d}$, the BER of the expurgated ensemble is upper bounded by

\begin{equation}
P_b\leq \frac{1}{\alpha}\mathop{\sum_{i=1}^{kN} \sum_{p=1}^{(n-k)N}}_{i+p\geq\hat{d}}\frac{i}{N} \bar{A}_{i,p}
Q\left ( \sqrt{2(i+p)R\frac{E_b}{N_0}}\right).
\end{equation}

\begin{figure}[t]
  \centering
    \includegraphics[width=\linewidth]{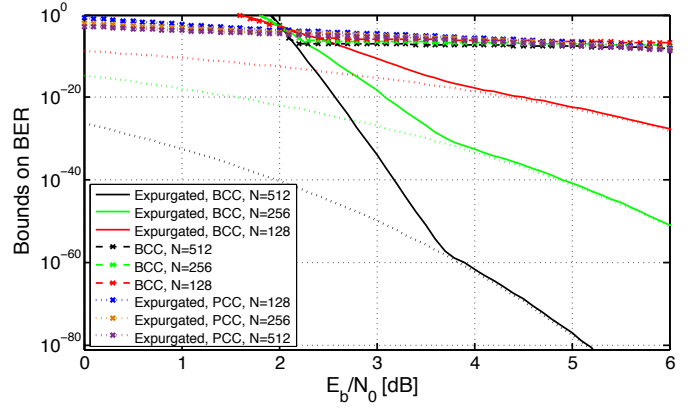}
\vspace{-2mm}
\caption{Expurgated union bound on the performance of PCC and BCC.}
\label{EXbounds}
\vspace{-3mm}
\end{figure}
For the BCC ensemble, the expurgated bounds on the BER are shown in
Fig.~\ref{EXbounds} for $\alpha=0.5$ and permutation sizes
$N=128, 256$, and $512$. The error floors estimated by the expurgated
bounds are much steeper than those given by the unexpurgated bounds. The
expurgated bounds on the BER are also shown in Fig.~\ref{EXbounds} for the PCC ensemble. These
bounds demonstrate that expurgation does not improve the performance
of the PCC ensemble significantly.

\section{Spatially Coupled \\ Braided Convolutional Codes}
\begin{figure}[!t]
  \centering
    \includegraphics[width=\linewidth]{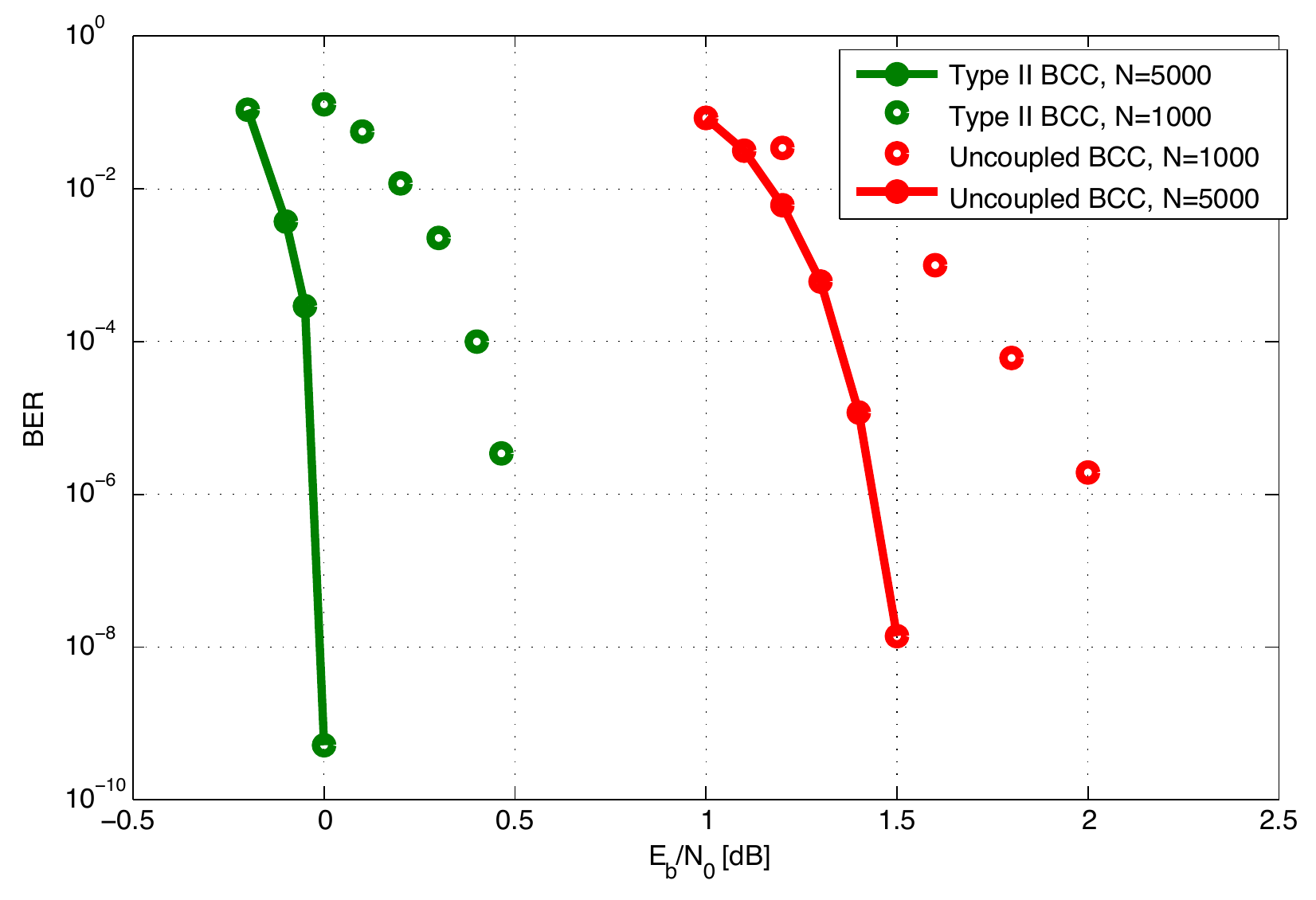}
\vspace{-3mm}
\caption{Simulation results for uncoupled and coupled BCCs with fixed permutations, $N=1000$ and $N=5000$.}
\label{SCBCCSim}
\vspace{-3mm}
\end{figure}

The performance of BCCs in the waterfall region can be significantly improved by spatial coupling. 
To demonstrate it, we provide simulation results for the uncoupled BCCs and Type-II BCCs for $N=1000$ and $5000$.
For coupled BCCs, we consider coupling length $L=100$ and a sliding window decoder with window size $W=5$ \cite{BCCWinDec}. 
For all cases, the permutations are selected randomly but fixed. 
Simulation results are shown in Fig.~\ref{SCBCCSim}. 
According to the figure, for a given permutation size, Type-II BCCs perform better than uncoupled BCCs.
As an example, for $N=5000$, the performance improves almost
$1.5\;\text{dB}$. 
We also compare the uncoupled and coupled BCCs with equal decoding latency.
In this case, we consider $N=5000$ and $N=1000$ for the uncoupled and coupled BCCs, respectively.
Considering equal decoding latency, the performance of the coupled Type-II BCCs is still significantly better than that of the uncoupled BCCs. 

The coupled BCCs have good performance in the waterfall region and their error floor is so low that it cannot be observed. 
It is possible to generalize equation \eqref{IPWEBCC} for the coupled BCC ensembles in Fig.~\ref{SCBCC} but the computational complexity is significantly increased.
In the following theorem, we establish a connection between the WEF of the uncoupled BCC ensemble and that of the coupled ensemble.
More specifically, we show that the weights of codewords cannot decrease by spatial coupling.
A similar property is shown for LDPC codes in \cite{TruhachevDistBoundsTBLDPCCCs,MitchellMinDisTrapSet2013,PseudocodLDPC}.

\begin{theorem}
Consider an uncoupled BCC $\tilde{\mathcal{C}}$ with permutations
$\Pi$, $\Pi^{\text{U}}$ and $\Pi^{\text{L}}$.  This code can be
obtained by means of tailbiting an original (coupled) BCC $\mathcal{C}$ with time-invariant
permutations $\Pi_t=\Pi$, $\Pi_t^{\text{U}}=\Pi^{\text{U}}$ and $\Pi_t^{\text{L}}=\Pi^{\text{L}}$. Let
$\bs{v}=(\bs{v}_1, \dots,\bs{v}_t, \dots, \bs{v}_L)$,
$\bs{v}_t=(\bs{u}_t,\bs{v}_t^{\text{U}},\bs{v}_t^{\text{U}})$, be
an arbitrary code sequence of $\mathcal{C}$. Then
there exists a codeword $\tilde{\bs{v}} \in \tilde{\mathcal{C}}$ that
satifies
\[
w_{\text{H}}(\tilde{\bs{v}}) \leq  w_{\text{H}}({\bs{v}}) \ , 
\]
i.e., the coupling does either preserve or increase the Hamming weight of
valid code sequences. 
\end{theorem}
\begin{IEEEproof}
A valid code sequence of $\mathcal{C}$ has to satisfy the local
constraints
\begin{align}
\begin{pmatrix}
\bs{u}_t & \bs{v}_{t-1}^{\text{L}} \cdot \Pi_t^{\text{U}} &
\bs{v}_{t}^{\text{U}} 
\end{pmatrix}
\cdot \bs{H}_{\text{U}}^T  & = \bs{0} \label{eq:coupledUpper} \\
\begin{pmatrix}
\bs{u}_t \cdot \Pi_t & \bs{v}_{t-1}^{\text{U}} \cdot \Pi_t^{\text{L}} &
\bs{v}_{t}^{\text{L}} 
\end{pmatrix}
\cdot \bs{H}_{\text{L}}^T  & = \bs{0} \label{eq:coupledLower} 
\end{align}
for all $t=1,\dots,L$, where $\bs{H}_{\text{U}}$ and
$\bs{H}_{\text{L}}$ are the parity-check matrices that represent the
contraints imposed by the trellises of the upper and lower component
encoders, respectively.  Since these constraints are linear and time-invariant, it follows
that any superposition of vectors
$\bs{v}_t=(\bs{u}_t,\bs{v}_t^{\text{U}},\bs{v}_t^{\text{U}})$ from
different time instants $t \in \{1,\dots,L\}$ will also satisfy
\eqref{eq:coupledUpper} and \eqref{eq:coupledLower}. In particular, if
we let
\[
\tilde{\bs{u}}=\sum_{t=1}^{L} \bs{u}_t \ , \quad
\tilde{\bs{v}}^{\text{L}}=\sum_{t=1}^{L} \bs{v}_t^{\text{L}}  \ , \quad \tilde{\bs{v}}^{\text{U}}=\sum_{t=1}^{L} \bs{v}_t^{\text{U}},
\]
then 
\begin{align}
\begin{pmatrix}
\tilde{\bs{u}} & \tilde{\bs{v}}^{\text{L}} \cdot \Pi^{\text{U}} &
\tilde{\bs{v}}^{\text{U}} 
\end{pmatrix}
\cdot \bs{H}_{\text{U}}^T  & = \bs{0} \label{eq:uncoupledUpper} \\
\begin{pmatrix}
\tilde{\bs{u}} \cdot \Pi & \tilde{\bs{v}}^{\text{U}} \cdot \Pi^{\text{L}} &
\tilde{\bs{v}}^{\text{L}} 
\end{pmatrix}
\cdot \bs{H}_{\text{L}}^T  & = \bs{0} \label{eq:uncoupledLower}
\enspace .
\end{align}
Here we have implicitly made use of the fact that $\bs{v}_t=\bs{0}$
for $t<1$ and $t>L$. But now it follows from \eqref{eq:uncoupledUpper}
and \eqref{eq:uncoupledLower} that $\tilde{\bs{v}}=(\tilde{\bs{u}},
\tilde{\bs{v}}^{\text{U}}, \tilde{\bs{v}}^{\text{L}}) \in
\tilde{\mathcal{C}}$, i.e., we obtain a codeword of the uncoupled
code. If all non-zero symbols within $\bs{v}_t$ occur at
different positions for $t=1,\dots,L$, then
$w_{\text{H}}(\tilde{\bs{v}}) =  w_{\text{H}}({\bs{v}})$. If, on the
other hand, the support of non-zero symbols overlaps, the weight of
$\tilde{\bs{v}}$ is reduced accordingly and  $w_{\text{H}}(\tilde{\bs{v}}) <  w_{\text{H}}({\bs{v}})$.
\end{IEEEproof}
\begin{corollary}
The minimum distance of the coupled BCC $\mathcal{C}$ is larger than
or equal to the minimum distance of the uncoupled BCC
$\tilde{\mathcal{C}}$,
\[
d_{\text{min}}(\mathcal{C}) \geq d_{\text{min}}(\tilde{\mathcal{C}}).
\] \hfill $\square$
\end{corollary}
From {\it{Corollary 1}}, we can conclude that the estimated
floor for the uncoupled BCC ensemble is also valid for the coupled BCC ensemble.

\section{Conclusion}

The finite block length analysis of BCCs performed in this paper, together with the DE analysis in \cite{JournalMLD}, show that BCCs are a very promising class of codes. They provide both close-to-capacity thresholds and very low error floors even for moderate block lengths.
However, we would like to remark that the bounds on the error floor in this paper assume a maximum likelihood decoder. In practice, the error floor of the BP decoder may be determined by absorbing sets. This can be observed, for example, for some ensembles of SC-LDPC codes \cite{Mitchell_AS2014}. Therefore, it would be interesting to analyze the absorbing sets of BCCs.

\end{document}